\documentclass{ws-procs975x65}

\def\beq{\begin{equation}}
\def\eeq{\end{equation}}
\newcommand{\pd}[2]{\frac{\partial #1}{\partial #2}} 

\begin{document}

\title{Spinning particles moving around black holes: integrability and chaos}
\author{Georgios Lukes-Gerakopoulos}
\address{Institute of Theoretical Physics, Faculty of Mathematics and Physics, 
 Charles University in Prague, Czech Republic\\
$^*$E-mail: gglukes@gmail.com}

\begin{abstract}
The motion of a stellar compact object around a supermassive black hole can be
approximated by the motion of a spinning test particle. The equations of motion
describing such systems are in general non-integrable, and therefore, chaotic
motion should be expected. This article discusses the integrability issue of the 
spinning particle for the cases of Schwarzschild and Kerr spacetime, and then
it focuses on a canonical Hamiltonian formalism where the spin of the particle
is included only up to the linear order. 
\end{abstract}

\keywords{Spinning test particle; black holes; Chaos.}

\bodymatter


\section{Introduction}

The dynamics of a spinning particle moving in a curved spacetime background is
of astrophysical interest as it approximates the motion of a stellar compact
object moving around a supermassive black hole. The equation of motion of a
spinning particle were provided by  Mathisson \cite{Mathisson37} and 
Papapetrou \cite{Papapetrou51}. However, the Mathisson-Papapetrou (MP) equations
are less than the variables they intend to evolve. Thus, a spin supplementary
condition (SSC) is required to close the system. There are aplenty SSCs
(see, e.g., Refs.~\citenum{3,4} for a review), but, here we mention 
the Tulczyjew\cite{Tulczyjew59} (T) SSC and the Newton-Wigner\cite{NewtonWigner49}
(NW) SSC. 

The dynamical reason we are interested in the spinning particle motion is that
chaos seems to appear for black hole backgrounds. Namely, it has been found that for
the MP equations with T SSC chaotic motion is present in the case of a Schwarzschild
background\cite{Suzuki97}, and in the case of a Kerr background\cite{Hartl03a,Hartl03b}.
However, it has been shown that when the MP equations with the T SSC are linearized,
then a Carter-like integral of motion appears in the case of the 
Kerr background\cite{Rudiger}, which led some to argue that in the linearized 
in spin approach the MP equation correspond to an integrable system.

A case of a linearized in spin system approach was employed in order to get a
canonical Hamiltonian formulation for the spinning particle system in
Ref.~\citenum{11}. Namely, this formulation came from linearizing in spin
the MP equations with the NW SSC. The Hamiltonian function for a Kerr background
in Boyer-Linquist coordinates given in Ref.~\citenum{11} suffers from several
drawbacks\cite{11,13}, and a revised Hamiltonian function in
Boyer-Linquist coordinates has been provided in Ref.~\citenum{12}. By
examining the latter function in the non-spinning limit of the central body,
i.e., Schwarzschild, we have found that the system is integrable, while for the
Kerr case chaos appears\cite{13}. The appearance of chaos will be examined
in detail in Ref.~\citenum{14}. In particular, in Ref.~\citenum{14} we
employ a 4D Poincar\'{e} method\cite{Patsis94,Katsanikas11a} to investigate the
dynamics of the spinning particle in the Hamiltonian approximation.

The article consists of the following sections. \Sref{sec:eqm} is a brief 
introduction to the canonical Hamiltonian formalism. \Sref{sec:integr} discusses
the integrability issue of a spinning particle in black hole background.
\Sref{sec:Conc} sums up the work.


\section{Equations of motion}\label{sec:eqm}

 A canonical Hamiltonian formalism has been achieved by linearizing the MP
 equations of motion for the NW SSC\cite{11}. The MP equations describe
 the motion of a particle with mass $m$ and spin $S^{\mu\nu}$ in a given
 spacetime background $g_{\mu\nu}$, i.e.,
 \begin{equation}
  \frac{D~p^\mu}{d \tau}=-\frac{1}{2}~{R^\mu}_{\nu\kappa\lambda}
  v^\nu S^{\kappa\lambda}~~,~
  \frac{D~S^{\mu\nu}}{d \tau}=p^\mu~v^\nu-v^\mu~p^\nu~~, \label{eq:MPeq}
 \end{equation}
 where ${R^\mu}_{\nu\kappa\lambda}$ is the Riemann tensor, $p^\mu$ is the 
 four-momentum, $v^\mu=d x^\mu/d \tau$ is the four-velocity, and $\tau$ is the
 proper time. The NW SSC reads 
 \begin{equation}
   S^{\mu\nu}~\omega_\mu=0~~, \label{eq:NW}
 \end{equation} 
 where $\omega_\mu$ is a sum of time-like vectors. This sum in our case\cite{11}
 has the form 
 \begin{equation}
  \omega_\nu=p_\nu-m~\tilde{e}_\nu^{~T}~~, \label{eq:NWTimeLV}
 \end{equation}
 where $\tilde{e}_\nu^{~T}$ is the timelike future oriented vector (T is used
 instead of 0), which together with three spacelike vectors
 $\tilde{e}^\mu_I$, is part of a tetrad field $\tilde{e}^\mu_\Delta$.

 When a tensor is projected on the tetrad field, then it is denoted with capital
 indices. For example, $\omega_\Delta=\tilde{e}^\nu_{~\Delta}\omega_\nu$ is the
 projection of the time-like vector (\ref{eq:NWTimeLV}) on the tetrad field, i.e.,
$
  \omega_T = p_\nu~\tilde{e}^\nu_{T}-m,~ \omega_J = p_\nu~\tilde{e}^\nu_{J}~.
$
 On the other hand, the spin tensor $S^{\mu\nu}$ projection reads
$
  S^{IJ}=S^{\mu\nu}~\tilde{e}_\mu^{~I}~\tilde{e}_\nu^{~J}~.
$
 However, the Hamiltonian function of the spinning particle\cite{11}
 does not use exactly the above described spin projection, instead employs the
 spin three vector 
$
  S_I=\frac{1}{2} \epsilon_{IJL}~S^{JL}~, 
$
 where $\epsilon_{IJL}$ is the Levi-Civita symbol.

 Now, the Hamiltonian function $H$ for a spinning particle 
 \begin{equation}
  H=H_{NS}+H_{S}~~, \label{eq:HamSP}
 \end{equation}
 splits in two parts. The first 
 \begin{equation}
  H_{NS}=\beta^{i}P_i+\alpha~\sqrt{m^2+\gamma^{ij}P_i P_j} \label{eq:HamNSP}
 \end{equation}
 is the Hamiltonian for a non-spinning particle, where 
 \begin{equation}
  \alpha = \frac{1}{\sqrt{-g^{00}}}~~,~
  \beta^i = \frac{g^{0i}}{g^{00}}~~,~ 
  \gamma^{ij} = g^{ij}- \frac{g^{0i}g^{0j}}{g^{00}}~~, 
 \end{equation}
 and $P_i$ are the canonical momenta conjugate to $x^{i}$ of the
 Hamiltonian~\eqref{eq:HamSP}, which can be calculated from the ``kinetic''
 momenta $p_i$ through the relation
 \begin{equation}
  P_i = p_i+E_{i\Gamma\Delta}S^{\Gamma\Delta}
  = p_i+\left(2 E_{iTJ}\frac{\omega_C}{\omega_T}+E_{iJC}\right)
 \epsilon^{JCL}~S_L~~,
  \label{eq:momentaHL}
 \end{equation}
 where
$
   E_{\nu\Gamma\Delta}=-\frac{1}{2}\left(g_{\kappa\lambda}~
   \tilde{e}^\kappa_{~\Gamma}
   ~\frac{\partial\tilde{e}^\lambda_{~\Delta}}{\partial x^\nu}
  + \tilde{e}^\kappa_{~\Gamma}~\Gamma_{\kappa\nu\lambda}~
    \tilde{e}^\lambda_{~\Delta}\right)
  \label{eq:EmuAB}
$
 is a tensor which is antisymmetric in the last two indices, i.e.,
 $E_{\nu\Gamma\Delta}=-E_{\nu\Delta\Gamma}$, and $\Gamma_{\kappa\nu\lambda}$ are
 the Christoffel symbols; the second part of the Hamiltonian
 \begin{equation}\label{eq:HamSPc}
   H_S=-\left(\beta^{i} F_i^C+F_0^C+\frac{\alpha~\gamma^{ij}P_i~F_j^C}
 {\sqrt{m^2+\gamma^{ij}P_i P_j}}\right) S_C 
  \end{equation}
 includes the spin of the particle, where 
$
  F_\mu^C=\left(2 E_{\mu TI}\frac{\bar{\omega}_J}{\bar{\omega}_T}+E_{\mu IJ}\right)
 \epsilon^{IJC}
$
 and
 \begin{eqnarray}
  \bar{\omega}_\Delta = \bar{\omega}_\nu~\tilde{e}^\nu_{~\Delta}~~,&
  \bar{\omega}_\nu = \bar{P}_\nu-m~\tilde{e}_\nu^{~T}~~,&
  \bar{P}_i = P_i~~,\nonumber\\
  \bar{P}_0 = -\beta^i~P_i-\alpha\sqrt{m^2+\gamma^{ij}P_i P_j}~~,& 
  \bar{\omega}_T = \bar{P}_\nu~\tilde{e}^\nu_{~T}-m~~,& 
  \bar{\omega}_J = \bar{P}_\nu~\tilde{e}^\nu_{~J}~~.\label{eq:omegabar}
 \end{eqnarray}

 The equations of motion for the canonical variables as a function of coordinate
 time $t$ read
 \begin{equation}
  \frac{d x^i}{dt} = \pd H{P_i}~~,
  \frac{d P_i}{dt} = -\pd H{x^i}~~,
 \frac{d S_I}{dt} = \epsilon_{IJC}\pd H{S_J}S^C~~\label{eq:HamEqMotion}~~.
 \end{equation}

 It should be noted that for the MP with NW SSC the mass $m=\sqrt{-p_\nu p^\nu}$
 is not a constant of motion\cite{11,LSK1}. However, in the procedure of
 linearizing in spin the MP equations to get the Hamiltonian approximation, 
 $m$ becomes a constant of motion\cite{11}. The explicit form of the
 revised Hamiltonian function for the Kerr spacetime in Boyer-Linquist coordinates
 can be found in Ref.~\citenum{12}; here the function is not presented,
 because it is lengthy.
 
 \section{The issue of integrability}\label{sec:integr}

 When the particle is non-spinning then \eref{eq:MPeq} gives the geodesic orbit,
 and the system is symplectic. In a symplectic system with each constant of
 motion the phase space can be reduced by two dimensions (one degree of freedom),
 while in a non-symplectic just by one dimension. In the case of a Kerr
 background for a geodesic orbit we have four integrals of motion, i.e., the
 mass of the particle, the energy, the component of the orbital angular momentum
 along the symmetry axis, and the Carter constant. The above integrals of motion
 are independent and in involution. For the geodesic motion we have 8
 dimensional phase space (four degrees of freedom), therefore, the system is
 integrable. 
 
 By including the spin to the evolution scheme, we increase the dimensions of
 the phase space, and we break in general its symplectic structure. Namely, we
 have 4 dimensions from the position $x^\mu$, 4 from the velocities $v^\mu$,
 4 from the momenta $p^\mu$, and 6 from the spin $S^{\mu\nu}$, so totally we 
 have 18 dimensions. In the Kerr case from the Killing vectors we get 2
 constants, but we  lose the  Carter constant. On the other hand in the
 Schwarzschild spacetime the Killing vectors provide 3 independent and in
 ivnolution constants. By including a SSC we get 3 independent constraints.
 For the MP with T SSC the measure of the particle's spin is preserved, and
 the measure of the spin as well, so we have 2 constants more. The preservation
 of the four velocity is the last constant, which gives us in total for the Kerr
 8 constants, while for the Schwarzschild 9 constants. But, since the system
 is not symplectic, the integrals are not enough to make the system integrable,
 and chaos appears both for the Schwarzschild\cite{Suzuki97} and for
 the Kerr\cite{Hartl03a,Hartl03b} background. 
 
 We need to linearize the MP equations in spin to decrease the dimensions, and
 to find new constants of motion. Namely, since the momenta and the velocities
 are parallel in the linear regime, the phase space decrease by 4, thus, we get
 a 14 dimensional phase space. A Carter like constant is regained for Kerr when 
 the MP equations with T SSC are linearized in spin\cite{Rudiger}. In the
 Schwarzschild limit the linearized in spin MP equations preserve the measure of
 the orbital angular momentum\cite{Apostolatos96} for the Pirani SSC, which in 
 the linear domain is the same with the T SSC. But, now the four-momentum 
 contraction gives the same constant with the four-velocity contraction, thus we
 have one integral less.
  
 In the Hamiltonian approximation the degrees of freedom are five, three come
 from the position variables, and two from the spin 3-vector. Thus, one needs
 five integrals of motion in order to have an integrable system. In the case of
 the Schwarzschild background we have the conservation of the total angular
 momentum giving two integrals, we have the conservation of the spin of the
 particle, since the system is autonomous the Hamiltonian function itself is a
 constant, and the measure of the orbital momentum is preserved as well\cite{13}.
 Thus, the Hamiltonian approximation corresponds to an integrable system in the 
 case of the Schwarzschild background\cite{13} in contrast with the non-integrability
 of the MP equations with T~SSC\cite{Suzuki97}.
 
 \begin{figure}
\begin{center}
\includegraphics[width=0.6\textwidth]{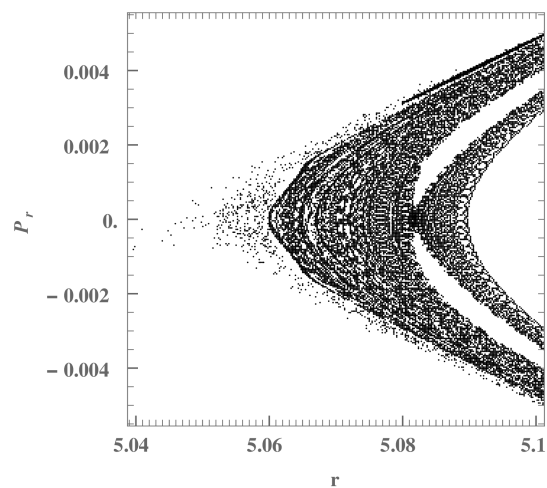}
\end{center}
\caption{A detail from the surface of section lying on the equatorial plane.
The basic parameters of the background spacetime are the mass $M=1$, and the 
spin $a=0.1$, while the particle's basic parameters are $S=m=1$. $r$ is the
radial distance in Boyer-Linquist coordinates, and $P_r$ the conjugate momentum
For more details see Ref.~\citenum{13}.
}
\label{fig:2D}
\end{figure}

 However, the Hamiltonian approximation for Kerr appears to correspond to a
 non-integrable system in agreement with the full MP equations for T~SSC\cite{Hartl03a,Hartl03b}.
 \Fref{fig:2D} shows a 2D surface of section, which is a projection of a 4D
 Poincar\'{e} section, for a slowly rotating central black hole $(a=0.1)$.
 The scattered dots in \Fref{fig:2D} show that chaos appears for the Hamiltonian
 function introduced in Ref.~\citenum{12}, which mean that the corresponding
 system is non-integrable. Moreover, the width of the KAM curves indicates that
 we have lost more than one integral when we turn on the spin of the central
 body. Since the azimuthal angular momentum, and the measure of the particle's
 spin are still constant of motion. The system has 3 degrees of freedom.
 The non-appearance of the Carter-like constant for the revised Hamiltonian
 function implies either that the tetrad field suggested in Ref.~\citenum{12}
 is not good, or that the Carter-like constant depends on the SSC we choose.
 
  \begin{figure}
\begin{center}
\includegraphics[width=0.45\textwidth]{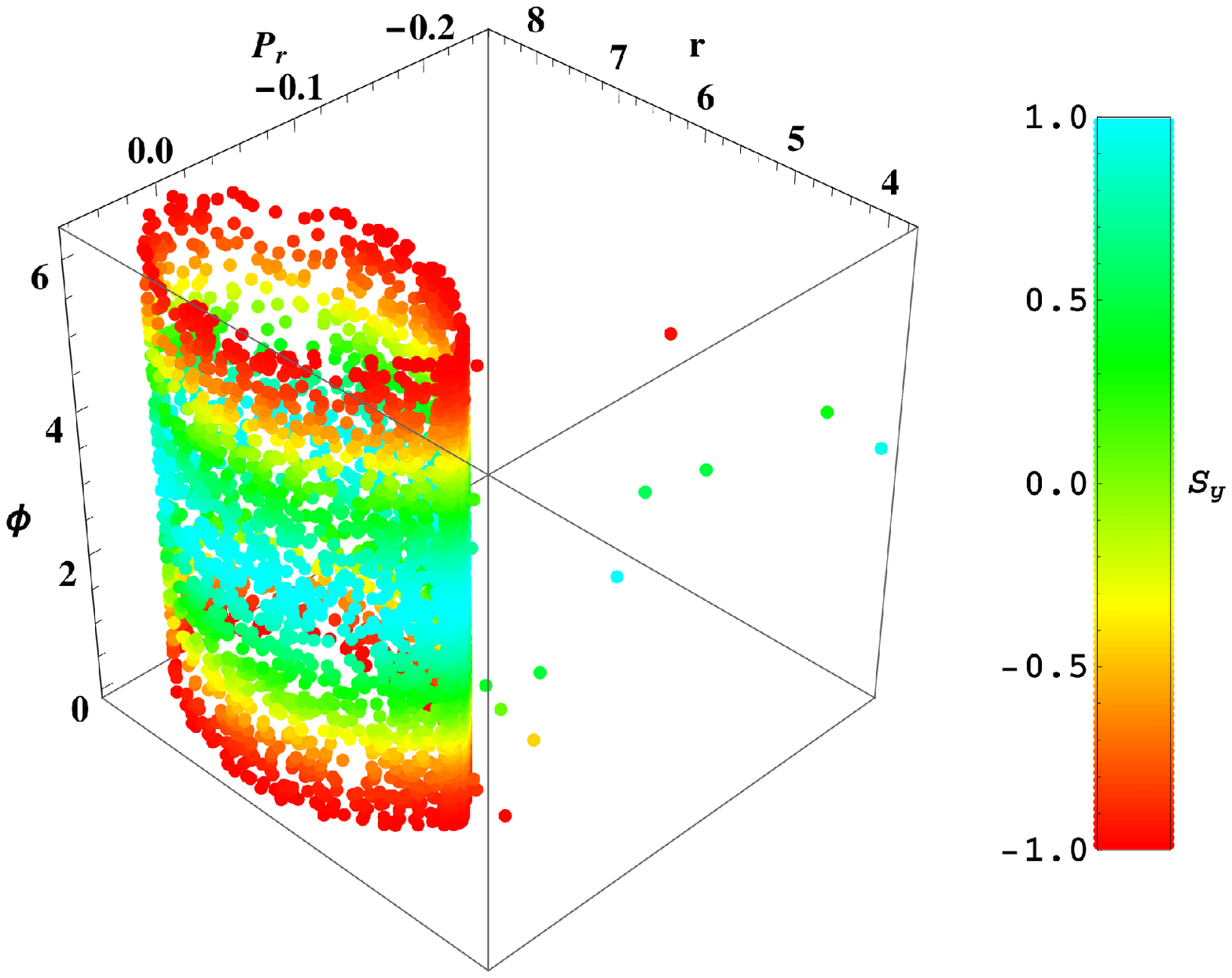}
\includegraphics[width=0.45\textwidth]{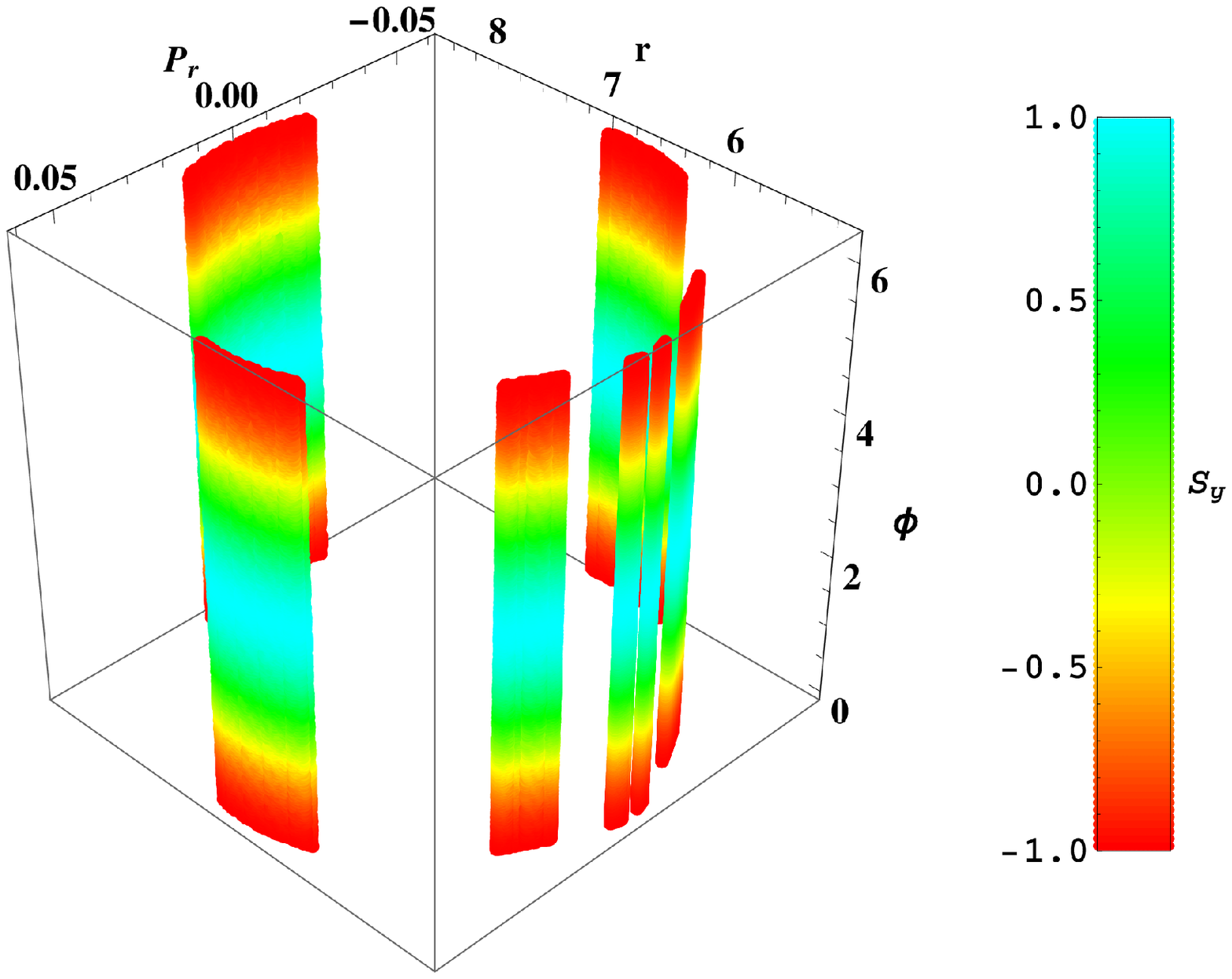}
\end{center}
\caption{A 4D Poincar\'{e} section corresponding to two orbits from \fref{fig:2D}.
$\phi$ is the azimuthal angle. $S_y$ is one of the three components of the 
spin, and the coloring corresponds to its value as shown at the strips lying 
right of each section. The left panel corresponds to sticky chaotic orbit shown
as scattered dots in \fref{fig:2D}, while the right panel corresponds to a chain
of stable islands corresponding to the two banana shaped white areas appearing
at the right sight of \fref{fig:2D}. 
}
\label{fig:4D}
\end{figure}
 
 The 2D Poincar\'{e} section method is proper for systems of two degrees of freedom,
 when the system has three degrees of freedom the 2D surface of section is just
 a projection of a 4D Poincar\'{e} sections. 2D projections are useful tool to
 get a first feeling about the dynamics and have been used in the
 past\cite{Suzuki97,Hartl03a,Hartl03b}. However, in order to confirm what we are 
 actually seeing, and to understand better the dynamics we need to use the
 4D Poincar\'{e} sections\cite{Patsis94,Katsanikas11a}. In \fref{fig:4D} we
 show some preliminary results from a work in progress\cite{14}. In this
 plot we present two 4D Poincar\'{e} sections, where the color plays the role
 of the fourth dimension. According to the color and rotation 4D Poincar\'{e}
 section method, the way an orbit evolves on a 3D projection along with the
 smoothness of the coloring shows whether an orbit is regular or 
 chaotic\cite{Patsis94,Katsanikas11a}.
 In the case of a chaotic orbit the orbit behaves irregularly on the 3D
 projection and/or the colors mix along its trajectory. In the case that a
 sticky chaotic orbit suffers from escapes like the case shown in \fref{fig:2D}
 the only way to observe the chaotic nature of the orbit is not the mixing of
 colors but the (irregular) scattered points leaving the sticky zone
 (left panel of \fref{fig:4D}). For an island of stability the
 regularity of the orbit is indeed shown by the toroidal structure of the 
 islands and the smoothness of the color (right panel of \fref{fig:4D}).
 
 \section{Conclusion}\label{sec:Conc} 
 
 This article has discussed the issue of integrability for the case of the 
 spinning particle, especially in the case when its dynamic is described by the
 canonical Hamiltonian linear in spin approximation. In the case of the
 Schwarzschild background, contrary to the non-integrability of the MP equations
 with T SSC\cite{Suzuki97}, the Hamiltonian approximation corresponds to an
 integrable system\cite{13}. While in the case of a Kerr background 
 the non-integrability of the MP equations with T SSC\cite{Hartl03a,Hartl03b} is
 also the case for the revised Hamiltonian function\cite{12}.
 The non-integrability of the latter is shown by a 2D surface of 
 section\cite{13}, and by using the color and rotation 4D Poincar\'{e} section
 technique\cite{14}. 

\section*{Acknowledgments}

G.L-G is supported by UNCE-204020 and by GACR-14-10625S.


\begin{thebibliography}{0}

 \bibitem{Mathisson37}
 M. Mathisson, {\em Acta Phys. Polonica} {\bf 6}, 163 (1937)
%
 \bibitem{Papapetrou51}
 A. Papapetrou, {\em Proc. R. Soc. London  Ser. A} {\bf 209},
 248 (1951)
%
 \bibitem{3}
 O. Semer\'{a}k, {\em Mon. Not. R. Astron. S.}
 \textbf{308}, 863 (1999)
%
 \bibitem{4}
  K. Kyrian, and O. Semer\'{a}k, {\em Mon. Not. R. Astron. S.}
 \textbf{382}, 1922 (2007)
%
 \bibitem{Tulczyjew59} W. Tulczyjew, {\em Acta Phys. Polonica}
\textbf{18}, 393 (1959)
%
 \bibitem{NewtonWigner49} T.~D. Newton and E.~P. Wigner,
 {\em Rev. Mod. Phys.} \textbf{21}, 400 (1949)
%
 \bibitem{Suzuki97} S. Suzuki and K. Maeda, {\em Phys. Rev. D}
 \textbf{55}, 4848 (1997)
%
 \bibitem{Hartl03a} M.~D. Hartl, {\em Phys. Rev. D} \textbf{67},
 024005 (2003)
%
 \bibitem{Hartl03b} M.~D. Hartl, {\em Phys. Rev. D} \textbf{67},
 104023 (2003)
%
 \bibitem{Rudiger} R.~R\"{u}diger, {\em Proc. R. Soc. London Ser.
 A} \textbf{375}, 185 (1981); \textbf{385}, 229 (1982) 
%
 \bibitem{11}
 E. Barausse, E. Racine, and A. Buonanno, {\em Phys. Rev. D} {\bf 80}, 104025 (2009)
%
\bibitem{12} E. Barausse, and A. Buonanno, {\em Phys. Rev. D} \textbf{81},
084024 (2010) 
%
 \bibitem{13}  D. Kunst, T. Ledvinka, G. Lukes-Gerakopoulos, and J. Seyrich, {\em Phys. Rev. D} {\bf 93}, 044004 (2016)
%
\bibitem{14}
G.~Lukes-Gerakopoulos, M.~Katsanikas, P.~Patsis
and J.~Seyrich, arXiv: 1606.09171, {\em Phys. Rev. D} (2016)
%
 \bibitem{Patsis94}
  P.~A. Patsis  and L. Zachilas {\em Int. J. Bif. Chaos} {\bf 4}, 
  1399-1424 (1994)
%
 \bibitem{Katsanikas11a}
  M. Katsanikas and  P.A. Patsis   {\em Int. Journal Bif. Chaos}  
  {\bf 21}, 467-496 (2011)
%
 \bibitem{LSK1} G. Lukes-Gerakopoulos, J. Seyrich, D. Kunst, 
 {\em Phys.Rev. D} \textbf{90}, 104019 (2014)
%
 \bibitem{Apostolatos96} T.~A.~Apostolatos, {\em Clas. Quant. Grav.}
 \textbf{13}, 799 (1996)
%
 \bibitem{Pirani56} F. A. E. Pirani, {\em Acta Phys. Polonica}
 \textbf{15}, 389 (1956) 

\end{thebibliography}
\end{document}